\documentclass{aa}  

\usepackage{graphicx}
\usepackage{tabularx}

\usepackage{txfonts}

\begin{document}

   \title{Deep chemical tagging}
   \subtitle{Identifying open clusters and moving groups in chemical space with graph attention networks}

   \author{Lorenzo Spina\inst{1, 2},
        Milan Quandt Rodriguez\inst{2, 3},
        Laura Magrini\inst{1},
        Leda Berni\inst{1,4},
        Sara Lucatello\inst{2},
        Marco Canducci\inst{5}
          }

   \institute{INAF - Osservatorio Astrofisico di Arcetri, Largo E. Fermi 5, 50125, Firenze, Italy --- \email{lorenzo.spina@inaf.it}\
         \and
             INAF - Padova Observatory, Vicolo dell’Osservatorio 5, 35122 Padova, Italy
        \and
        Dipartimento di Fisica e Astronomia, Universit\'a di Padova,I-35122, Padova, Italy
        \and
        Dipartimento di Fisica e Astronomia, Università degli Studi di Firenze, Via Sansone 1, 50019, Sesto Fiorentino, Italy
        \and
        University of Birmingham, School of computer science, B15 2TT, Birmingham, United Kingdom             }

   \date{Received June 3, 2025; accepted August 11, 2025}

  \abstract
   {Reconstructing the formation history of the Milky Way is hindered by stellar migration, which erases kinematic birth signatures. In contrast, stellar chemical abundances remain stable and can be used to trace stars back to their birth environments through chemical tagging.}
   {This study aims to improve chemical tagging by developing a method that leverages kinematic and age information to enhance clustering in chemical space, while remaining grounded in chemistry.}
   {We implement a graph attention auto-encoder that encodes stars as nodes with chemical features and connects them via edges based on orbital similarity and age. The network learns an ``informed'' chemical space that accentuates coherent groupings.}
   {Applied to $\sim$47,000 APOGEE thin disk stars, the method identifies 282 stellar groups. Among them, five out of six open clusters are successfully recovered. Other groups align with the known moving groups Arch/Hat, Sirius, Hyades, and Hercules.}
   {Our approach enables chemically grounded yet kinematically and age informed chemical tagging. It significantly improves the identification of coherent stellar populations, offering a framework for future large-scale stellar archaeology efforts.}


   \maketitle

\section{Introduction}

Understanding the structure and formation history of the Milky Way is one of the main challenges in astrophysics for the current decade. A major obstacle in unraveling the Galaxy’s evolution lies in the complex interdependencies among main stellar properties, such as spatial distribution, kinematics, chemical abundances, and ages. In particular, it is well established that stars can migrate and disperse throughout the Galactic disk \citep[e.g.,][]{Sellwood02,Rokach08,Schonrich09}. As a result, after a few interactions with the Galaxy’s gravitational potential, stars can lose all kinematic information related to their birthplaces. This loss severely hampers our ability to reconstruct the evolution of the Milky Way in space and time.

Contrary to kinematics, the chemical composition of stellar atmospheres -- with the exception of light elements -- remains largely unchanged over their lifetimes. This opens the possibility of using stellar chemical patterns to trace stars back to their birth environments and recover dispersed stellar populations such as open clusters, associations, or even moving groups, provided they share a common origin \citep{Freeman02}. The approach of identifying stars that originated from the same formation site based on their chemical abundances is known as ``chemical tagging.''

In recent years, several studies have tested the feasibility of chemical tagging within the thin disk. These efforts have mainly focused on applying unsupervised clustering algorithms directly in chemical abundance space, with the goal of blindly recovering known open cluster members \citep[e.g.,][]{Mitschang13,BlancoCuaresma15, Hogg16, Kos18, Casamiquela21}. Although some works have shown that methodological choices -- such as careful feature engineering, pre-processing, and the adoption of more suitable distance metrics -- can lead to some improvements in clustering performance \citep[e.g.,][]{GarciaDias19,Spina22}, the overall results have been limited. Even though there is growing evidence that the information about a star’s birth environment, its age, and its chemical abundance pattern are tightly linked \citep[e.g.,][]{Jofre17, Ness19, Ratcliffe23, Plotnikova24, Signor24,Wang24}, chemical tagging in the thin disk remains a particularly challenging task. The primary limitation stems from the high level of chemical homogeneity in the thin disk \citep[e.g.,][]{Ness18, Ting21,PriceJones20,Bhattarai24}, which significantly reduces the contrast between co-natal and unrelated stars in abundance space, ultimately hindering our ability to robustly reconstruct disrupted stellar populations. 

Crucially, the effectiveness of chemical tagging is tightly linked to the precision with which chemical abundances are measured: even subtle abundance differences can be meaningful if the observational uncertainties are sufficiently small \citep[e.g.,][]{Ting15}. Therefore, the success of chemical tagging depends not only on the intrinsic chemical diversity of the disk, but also on the capability of spectroscopic surveys to achieve the necessary abundance precision across a wide range of elements.

Given that current and upcoming spectroscopic surveys still fall short of delivering the abundance precision required for fully effective chemical tagging \citep{Mead25} some studies have explored alternative approaches such as ``chemodynamical tagging.'' This method combines chemical information with kinematic or orbital parameters to identify co-natal stellar groups \citep[e.g.,][]{Barth25}. While this strategy can improve clustering performance, especially when purely chemical signatures are insufficiently distinctive, it risks shifting the focus away from chemistry altogether. In practice, chemodynamical tagging ends up relying predominantly on orbital similarity, becoming more akin to classical dynamical membership analyses of open clusters. In such cases, the added value of chemical information is diminished, and the approach may no longer represent a true test of the chemical tagging paradigm. 

Nonetheless, kinematic information does exist and may provide valuable auxiliary constraints when used appropriately. While an overreliance on orbital parameters can obscure the chemical dimension of tagging, incorporating kinematics -- without letting it dominate -- can help mitigate some limitations of chemical-only approaches. A similar argument applies even more to stellar ages. For certain classes of evolved stars, particularly red giants with metallicities in the typical range found in the disk, ages can even be inferred directly from abundance ratios such as [C/N], which evolve in a predictable manner due to stellar nucleosynthesis during the red giant phase \citep[e.g.,][]{Martig16, Casali19, Anders23}. Therefore, chemical tagging approaches can and should benefit from ages derived through these so-called ``chemical clocks.''

Taking inspiration from the recent study by \citet{Berni25}, in this paper, we aim to develop and test a new analysis method designed to overcome some of the limitations that currently hinder the applicability of chemical tagging, particularly in the context of the low precision of spectroscopic surveys. 
Crucially, our method remains chemically grounded: the clustering is carried out in a chemical space that is selectively informed by stellar kinematics and ages, whenever these prove chemically relevant. This approach preserves the core philosophy of chemical tagging, while allowing for a more flexible and data-driven representation of abundance space. This framework offers a new pathway for improving our ability to identify dispersed stellar groups in the thin disk without compromising the chemical basis of the method.

In section~\ref{Sec:Methods} we describe the new method we develop for the analysis. section~\ref{Sec:Dataset_processing} gives details on the dataset we use and how these data are preprocessed. The analysis itself is described in section~\ref{Sec:Analysis}. The results are presented in section~\ref{Sec:results} and concluding remarks are given in section~\ref{Sec:conclusions}.

\section{Methods}
\label{Sec:Methods}

The goal of this work is to develop and test an innovative analytical approach that enables a synergistic and efficient use of both kinematic and chemical information. Nonetheless, the method remains fundamentally anchored in chemistry, thereby preserving the core principles and foundations of chemical tagging. 

A general underlying assumption of chemical tagging is that while stars can migrate and change their kinematical properties, their chemical pattern is invariant and depends on the environment in which they were born. Hence, our objective is using chemistry to determine whether or not kinematics retains relevant information on the origin of a star. Then, in such cases, we take advantage of kinematics to generate an ``informed'' representation of the stellar chemical pattern, which is more suitable for chemical tagging. This method of analysis is built upon the use of a graph attention auto-encoder (GATE). 

Auto-encoders are specific types of neural networks used for unsupervised learning, designed to encode data into compressed representations and then reconstruct it as accurately as possible. They consist of three main components: the encoder, the bottleneck, and the decoder. The encoder is a series of layers with a progressively smaller number of neurons. As the input data passes through these layers, it is compressed by being ``squeezed'' into a lower-dimensional space, determined by the number of neurons in the last layer of the encoder. The bottleneck is this smallest and most compressed layer in the network. This is both the output of the encoder and the input of the decoder. The bottleneck layer is an essential part of the auto-encoder as it creates the so-called latent space: a reduced-dimensional representation of the input data, which captures the most essential features. Once the information is compressed to pass through the bottleneck, the decoder takes this latent representation and reconstructs the original data. The decoder's architecture mirrors that of the encoder, with the same number of layers but a progressively increasing number of neurons. Thanks to this structure, the decoder can expand the data back to its original dimensions. Due to their ability to create a latent space, auto-encoders are typically used to reduce the dimensionality of data while preserving its structure and the most important features (e.g., proximity of similar data points, clusters, continuity, correlations). However, these neural nets could also be used to reduce the level of noise in data. In fact, if noise cannot be learned and represented as a pattern in the latent space, then the decoded representation of the data should be a cleaner version of the original input, with the noise filtered out.

\begin{figure}
\centering
\includegraphics[width=\hsize]{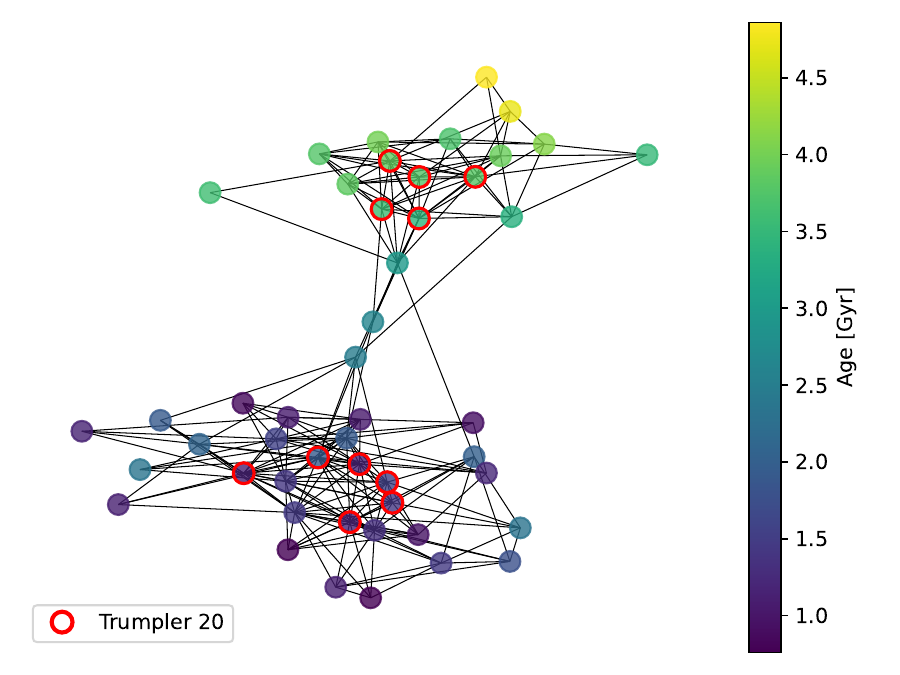}
    \caption{Portion of the graph processed by the GATE . This subgraph shows all nodes belonging to Trumpler~20 and all their immediate neighbors. Edges between all nodes are shown as solid black lines. Cluster members are highlighted by a red circle. Each node is color-coded based on the stellar age. Nodes are arranged in the plane to minimize the distance between connected nodes.}
        \label{Fig:Trumpler_20_graph}
\end{figure}

Real-world objects are often defined not only by their individual features but also by their mutual connections. This is especially true for complex systems, populations, or, more generally, entities with interactions. A dataset that includes both features and connections can be represented as a ``graph''. An illustrative example of a graph is shown in Fig.~\ref{Fig:Trumpler_20_graph}. A graph is a structure that visualizes a collection of data points, known as ``nodes,'' and their mutual connections, called ``edges''. Typically, edges are directional, indicating that each node may have connections either incoming or outgoing. When a connection between two nodes goes in both directions, the edge is considered ``undirected''. In addition to classical edges bridging two different nodes in a graph, nodes can also have ``self-loops,'' meaning that they have edges pointing to themselves. Unlike traditional tabular datasets, where each data point is independent and identically distributed, graphs inherently capture relational information. In a graph, the relationships between entities are as crucial as the entities themselves. For instance, the graph in Fig.~\ref{Fig:Trumpler_20_graph} is constructed so that two nodes with similar ages are connected by an edge.

Graph neural nets (GNNs) are a class of artificial neural networks specifically designed to operate on graph-structured data \citep{kipf2017}. While basic neural networks perform only on node features by considering them as tabular datasets, GNNs enable the integration of both node and edge information into the learning process (see \citealt{Berni25} for an application of GNN to stellar populations). While a basic neural network layer corresponds to a linear tranformation H~=~XW$^{T}$, where X is the input matrix and W is the weight matrix, the graph linear layer can be rewritten as H~=~A$^{T}$XW$^{T}$. Here, A is a square matrix where each row and column correspond to a node in the graph and where values -- typically either 0 or 1 -- indicate whether there is an edge between two nodes. Thus, at each layer of the GNN, information from each node is aggregated and combined with that of all nodes connected to it. The aggregation is also weighted by the number of edges connected to the node. That is done to allow a balance between the case of a node with many edges and a node with few or no edges, so that nodes with more edges do not have a higher overall influence on the neuron's output. A graph attention network (GAT) is a further development of GNNs that assigns a weight to each edge, representing the importance of a neighboring node’s features when updating a given node’s representation \citep{velickovic2018}. This weight, denoted as $\omega$, determines the influence of each connection in the graph. In practice, the graph attention linear layer is H~=~A$^{T}$W$_{\omega}$XW$^{T}$, where W$_{\omega}$ stores the weights for all edges in the graph. These weights, which are learnable parameters, adjust the contribution of each neighboring node to the updated node representation. By design, the weights assigned to incoming edges for each node (including the self-loop) are normalized so that their total sum equals one.

Our technique is based on an auto-encoder built using a GAT. The nodes' features are the chemical abundances derived for the stars in our sample, meaning that the auto-encoder will aim at reproducing these abundances. However, stars with similar orbital actions and similar ages are connected by edges, implying that the GAT's nature of the auto-encoder progressively learns the importance of each link for accomplishing the task. As we will show in section~\ref{Sec:Analysis}, the GATE will naturally assign most of the weight to the self-loops, as they are the most important edge for each node to reconstruct its input features. Hence, any edge connecting two different nodes will be effectively removed, with its final weight approaching zero. However, during training, the GATE first eliminates edges between stars with significantly different chemical patterns. Then, it progressively breaks connections between stars with increasingly similar chemistry until, eventually, only self-loops remain. 

This behavior highlights the crucial role of self-loops in the learning process. In particular, for stars connected to chemically dissimilar neighbors, self-loops are essential to preserve their intrinsic chemical information during reconstruction. Without a strong self-loop, a node would aggregate and prioritize the features of unrelated neighbors, leading to a distorted chemical signature. Additionally, self-loops provide a consistent reference scale that enables the normalization of edge weights across the graph. By comparing the weight of each edge to that of the self-loop, we obtain a measure of relative similarity between neighbouring nodes that is both interpretable and comparable across the whole graph.

Our goal is then to stop the training process at an epoch where the edges between very similar nodes are still important, while all the others already have small weights. We determine this sweet point along the training process by following the progressive adjustment of the $\omega$ weights between stellar members of some open clusters in our dataset and compare it to those of other random edges. 

At the end of the entire process we gather the following information;

\begin{itemize}
    \item Information on edges. Stars that have removed most of their edges are presumably those that have significantly migrated from their native region within the Galaxy. In fact, for these stars the neural network recognizes that their connections to stars with similar kinematics and ages has a negative influence in predicting its actual chemical pattern. On the other hand, stars that have retained many of their edges are presumably those that have not migrated significantly or those within clusters and associations.
    
    \item Information on GATE's output. The GATE is trained to predict the best representation of the stellar chemical patterns. However, each neuron aggregates the input features from connected stars. Hence, a group of stars fully connected by edges results much closer to each other in the ``informed'' chemical space than what they are in the original space. As a result, groups of stars with similar chemical patterns, kinematics and ages can be identified as high-density clumps in the chemical space predicted by the GATE even if they were much sparser in the original space. 
    
\end{itemize}

This paper mainly focuses on this latter piece of information. Namely, here we show that a clustering analysis of the chemical space predicted by the GATE allows the efficient recovery of open clusters with only 10-30 members hidden within a huge dataset of $\sim$40k thin disk stars. A following paper will instead focus on the first piece of information, aiming at understanding how stars migrate within the Galactic disk.

\section{Dataset and processing}
\label{Sec:Dataset_processing}

The method of analysis described in the previous section makes use of three different types of data. First is the chemical abundances, which define the features of the nodes. Then, we also require astrometric information and radial velocities from which we derive the orbital parameters. Finally, the stellar ages, which -- together with the orbital parameters -- allow us to establish the edges between the nodes. In this section we describe the initial steps of the analysis, from the choice of the dataset to the creation of the graph, including the identification of the open clusters' members that we use to test and validate our method.

Given these premises we choose the same APOGEE DR17 dataset \citep{Abdurrouf22} for which \citet{Anders23} has provided stellar ages. We clean this dataset by removing all the instances with STARFLAGs equal to BAD\_PIXELS, BRIGHT\_NEIGHBOR, VERY\_BRIGHT\_NEIGHBOR, LOW\_SNR, PERSIST\_HIGH, PERSIST\_MED, PERSIST\_LOW, PERSIST\_JUMP\_POS, PERSIST\_JUMP\_NEG, SUSPECT\_BROAD\_LINES, MULTIPLE\_SUSPECT, SUSPECT\_ROTATION, MTPFLUX\_LT\_75, and MTPFLUX\_LT\_50. We also remove all stars with signal-to-noise ratio below 100 pxl$^{-1}$, ASPCAP bitmasks greater than 16, microturbolence above 2~km~s$^{\rm-1}$ and below 0.75~km~s$^{\rm-1}$. Given that we are only interested in thin disk stars, we remove all stars with ages greater than 9 Gyr. However, since the resulting dataset still contain a number of ``chemically'' thick disk stars, we carefully inspected the [$\alpha$/M]-[M/H] diagram and rejected all the stars located above the thin disk sequence. After this cleaning procedure there are 61,564 stars remaining.

After this first step of data cleaning we identify the members of open clusters within the dataset. For this aim we use the all-sky cluster catalog from \citet{Hunt23}, selecting all stars matching with at least one object in our dataset, with cluster's membership probability greater than 0.5 and located within the cluster's tidal radius. Given these criteria, then we select all clusters with at least three members and we perform a 2-sigma clipping in the radial velocity from APOGEE in order to remove false members. The clipping was performed in three iterations using the median as the center value. As a result, we identify 14 open clusters with three to nine members. These members are used throughout the analysis for driving our decisions, such as the chemical elements to be used or the epoch at which the GATE's learning should be stopped. In addition to these, we identify seven additional clusters with ten or more members which are those that we want to recover with this method and are used to evaluate the full analysis.

In order to define which are the most meaningful chemical elements for our analysis, we follow the same procedure described in \citet{Spina22}. We use the members from the 14 open clusters to calculate the Calinski-Harabasz score for each chemical element detected by APOGEE and we select for the further analysis only those with score higher than 5, with the only exception of Mg. In fact, the MG\_FE abundance ratios are strongly correlated with ALPHA\_M values (i.e., correlation coefficient equal to 0.93). Since these latter have a larger Calinski-Harabasz score, the ALPHA\_M abundance ratios are preferred over MG\_FE. Given all these criteria, we select ten chemical abundances for the analysis: M\_H, ALPHA\_M, AL\_FE, K\_FE, CA\_FE, TI\_FE, V\_FE, MN\_FE, NI\_FE, CE\_FE.

Once the most meaningful elements are identified, we further clean the dataset by removing all stars whose abundances are determined with poor precision. To do so, we only keep the stars with uncertainties in all these abundances that are smaller than their 85$^{th}$ percentiles. That leaves the dataset to 47,346 stars. 

For these stars we derive the orbital actions using the same method described in \citet{Spina21}. Namely, we make use of the stellar sky-coordinates and proper motions from Gaia DR3, distances are from \citet{BailerJones21}, and radial velocities from APOGEE DR17. Given a star, for each of these quantities we draw 10,000 samples from normal distributions centred on their nominal values and a standard deviation equal to their uncertainties. For each of this samples we calculate with GALPY \citep{Bovy15} the orbital actions L$_{z}$, J$_{r}$, and J$_{z}$. The final values used in this work for that single star are the median of the resulting distributions. Through the same procedure we also derive the Galactocentric distance (R$_{\rm Gal}$) and cylindrical velocities (V$_{\rm \phi}$, V$_{\rm R}$, V$_{\rm z}$). Once the orbital parameters have been obtained, we further remove around 50 stars having extreme orbital actions, namely L$_{z}$ $<$ -200~kpc~km~s$^{-1}$, L$_{z}$ $>$ 3500~kpc~km~s$^{-1}$, $\sqrt{J_{z}}$ $>$ 20~kpc~km~s$^{-1}$, and $\sqrt{J_{r}}$ $>$ 20~kpc~km~s$^{-1}$. These are stars with large uncertainties in the astrometric parameters or radial velocities or contaminants belonging to the Galactic halo. After this final cleaning the dataset contains 47,281. These are the nodes of the graph that we process through the GATE. Each of these nodes has ten features, which are the chemical abundances identified above.

The edges of the graph are instead determined from the same orbital actions derived as above and ages from \citet{Anders23}. The incoming edges of a single star are defined by its ten nearest neighbors in the standardized orbital actions space, provided their ages differ by no more than 1 Gyr. Most of the edges established through this criterion are undirected. However, also those going only in one direction have been transformed to undirected. As a result, each node has on average 12 edges. 

The final graph contains 47,281 nodes and 578,110 edges. It includes ten open clusters with three to nine members and six open clusters with ten members or more. The open clusters with fewer than ten members have been used to drive some important decisions in our analysis, such as the model selection. This subset counts 45 stars and it is called calibration set. Instead, the open clusters with ten or more members are those that we want to retrieve through our analysis: NGC~7789 (29 stars), NGC~6819 (20), NGC~2682 (13), NGC~188 (11), Trumpler~20 (11), and NGC~6705 (10) This subset, which is called the ``target set,'' is never used any point during model design or calibration. Therefore, the retrieval performances of these six clusters -- discussed in section~\ref{Sec:open_clusters} -- provide empirical validation for the generalization capability of out method. 

As an example, Fig.~\ref{Fig:Trumpler_20_graph} we show the subgraph extracted around the members of Trumpler~20, which is one of the target clusters. Here we are deliberately showing an atypical case to illustrate how kinematics and ages model the structure of a graph. In fact, although the isochronal age of this cluster is 2 Gyr \citep{Cavallo24}, its members have ages from \citet{Anders23} clumping at 1.4 and 4 Gyr. This dichotomy creates a graph with two densely connected components and sparse interconnections. 

\section{Analysis}
\label{Sec:Analysis}

In this section we describe the procedure of analysis. The first step is training the GATE (section~\ref{Sec:training}). During this operation models for each epoch are saved. The second step consists in selecting the model to be used to generate the informed chemical space (section~\ref{Sec:model_selection}). The third step consists in the model's validation, which is carried out by showing that the predicted edges' weights correctly capture information on the chemical similarity between stars (section~\ref{Sec:validation}). Finally, the fourth step is the clustering analysis and the identification of groups of stars having a similar chemical make-up (section~\ref{Sec:clustering}).

\subsection{Training the GATE}
\label{Sec:training}

The GATE is composed of five single-head graph attention layers of 50, 50, 4, 50, and 50 neurons each. The four dimensions of the latent space represent the three main channels of nucleosynthesis of the chemical elements considered here (i.e., core-collapse supernovae, type II supernovae, asymptotic giant branch stars) plus an additional dimension that would account for an extra degree of diversity between the production sites. More complex network architectures have been tested with no or small improvement in the reconstruction loss. Hence, we opted for the relatively simple structure described above, which provides an optimal balance between model complexity and performance. All network's layers are followed by LeakyReLU activation functions with a 0.2 slope, with the only exception of the last layer which is connected to a linear activation function. The GATE is coded using the \texttt{GATv2} operator from \texttt{torch\_geometric}. The optimizer is \texttt{AdamW} with a learning rate of 0.01. However, we also use a learning rate scheduler with a 0.5 factor and patience of 10. In order to ensure a more efficient training of the GATE, we standardize the ten nodes' features. Then, 10$\%$ of graph's nodes are used for validation, while the remaining 90$\%$ for the actual training. The GATE is trained with mini batches of 1000 nodes uniformly distributed in the [M/H]-age space. The GATE is trained over a main squared error loss and for 150 epochs.

\subsection{Model selection}
\label{Sec:model_selection}

During training the GATE will necessarily learn that the most important edge is the self-loop. For that reason it will progressively remove weight to all the other edges starting from those between stars with extremely different chemical patterns. The main goal of this analysis' step is identifying the training epoch -- and the corresponding model -- at which edges between chemically different stars have very little weights, but not all the other edges. 

\begin{figure}
\centering
\includegraphics[width=\hsize]{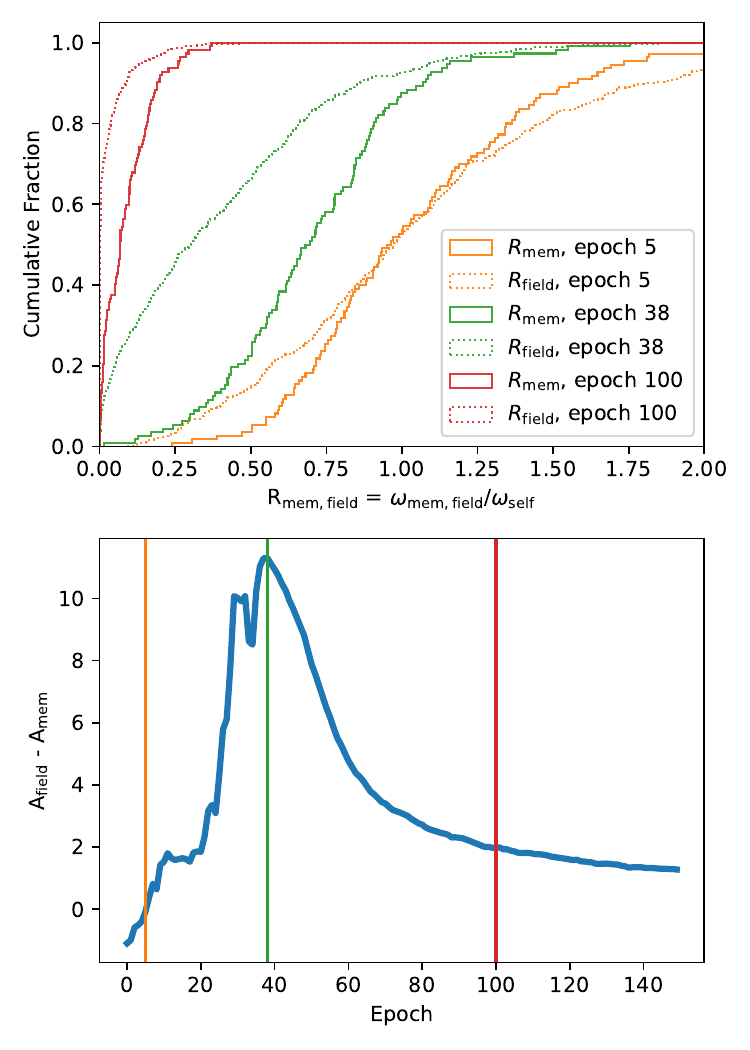}
    \caption{\textbf{Top panel}. Cumulative distributions of R$_{\rm mem}$ and R$_{\rm field}$ derived for the epochs 5, 38, and 100. \textbf{Bottom panel}. Difference between the area of the cumulative distributions calculated for R$_{\rm field}$ and that obtained for R$_{\rm mem}$, as a function of the training epoch. The orange, green and red vertical lines correspond to the distributions shown in the top panel. The peak of this curve identifies the model that better captures the chemical similarity/diversity between connected stars.}
        \label{Fig:model_selection}
\end{figure}

To do so, we consider the 45 stars of the calibration set. These are known members of open clusters. Each one of these stars is linked to both other members of their own open cluster and field stars. These stars from the calibration set form 112 edges between members of the same open cluster. Their weights are written as $\omega_{\rm mem}$. They also form 483 edges between stars that are not belonging to the same cluster. Their weights are written as $\omega_{\rm field}$. For both these classes of edges we derive the ratios R$_{\rm mem}$ = $\omega_{\rm mem}$/$\omega_{\rm self}$ and R$_{\rm field}$ = $\omega_{\rm field}$/$\omega_{\rm self}$, where $\omega_{\rm self}$ is the weight of the self-loop of the corresponding node. In Fig.~\ref{Fig:model_selection}-top we show the cumulative distributions of R$_{\rm mem}$ and R$_{\rm field}$ derived for the epochs 5, 38, and 100. At epoch 5 the GATE has still not learnt much from data, thus $\omega_{\rm mem}$ and $\omega_{\rm field}$ are still similar to their initial values and to their $\omega_{\rm self}$. Instead, at epoch 38 the R$_{\rm field}$ span lower values than R$_{\rm mem}$, meaning that $\omega_{\rm field}$ have been lowered during training much more than $\omega_{\rm mem}$. Finally, at epoch 100 we see that both R$_{\rm field}$ and R$_{\rm mem}$ are very low, meaning that the self-loops have already acquired most of the total weight at the expenses of all the other edges, regardless if they are between stars belonging to the same cluster or not. 

The model we wish to use among the 150 generated during the training is the one that maximizes the difference between the area under the cumulative distributions of R$_{\rm mem}$ and R$_{\rm field}$. In the bottom panel of Fig.~\ref{Fig:model_selection} we show these areas as a function of the epoch. The maximum difference between these areas occurs at epoch 38. Its model (i.e., \texttt{model$\_$38}) better reflects the chemical diversity between connected stars in the edges' weights. That is the model that we use for the following analysis.

To test the robustness of this analysis, we repeated the model selection across all possible combinations of five clusters out of the ten available in the calibration set, which amounts to 252 cases. The optimal epoch varies by 30–40 and systematically shifts earlier when lower-S/R spectra are considered. In fact, increased noise spreads cluster members in chemical space, making early epochs -- where dissimilar stars remain strongly connected -- more informative.

\subsection{Validating \texttt{model$\_$38}: Migration of stars}
\label{Sec:validation}

In general, stars that have retained most of the weights in their inward edges (i.e., $\omega$) have chemical patterns that are similar to other coeval stars with similar orbital actions. These stars have likely not undergone significant orbital changes. On the other hand, stars whose chemical patterns are strongly dissimilar to coeval neighbors in the orbital actions' space should have seen their inward weights drastically reduced. These stars have likely undergone strong orbital perturbations, to the extent that their present-day orbital actions bears little or no relation to their original birth location within the Galaxy. To evaluate whether \texttt{model$\_$38} effectively captures the chemical diversity of stars, we test whether these predicted relationships between chemical similarity and orbital parameters hold in practice.

\begin{figure}
\centering
\includegraphics[width=\hsize]{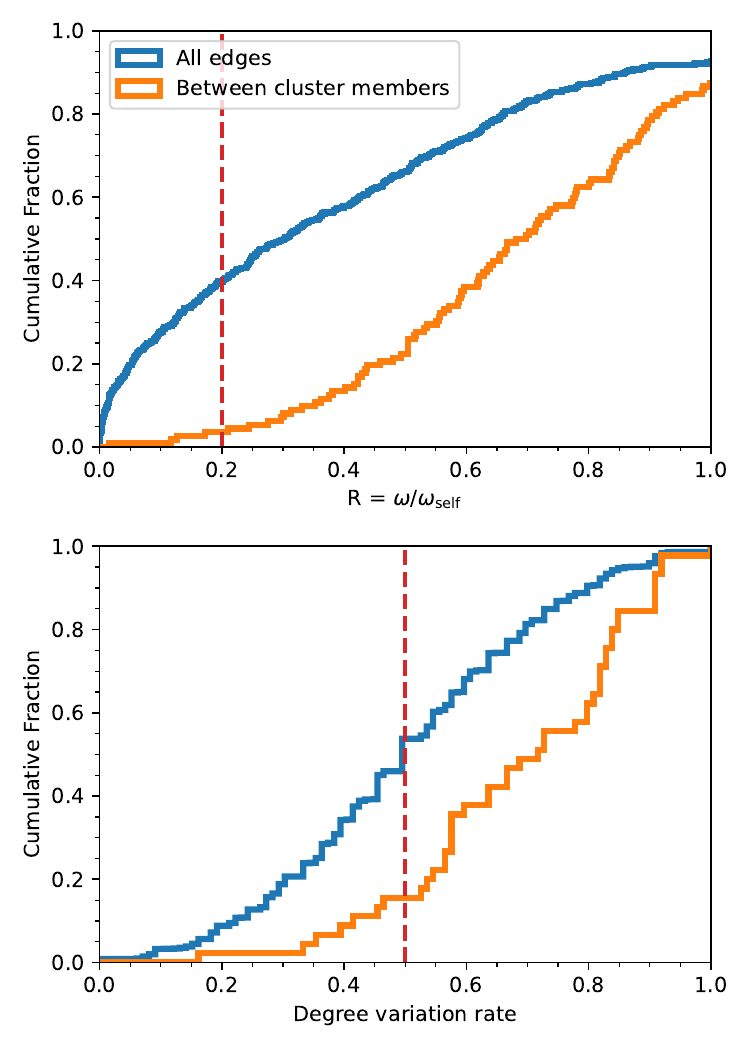}
    \caption{\textbf{Top panel}. The histogram shows the cumulative fraction of the ratio R = $\omega$/$\omega_{\rm self}$ for all edges (blue) and for the edges connecting members of the same cluster from the calibration set (orange). The dashed red line indicates the threshold below which we consider the edge as non-significative and we remove it from the graph. \textbf{Bottom panel}. The histogram shows the cumulative fraction of the ratio between the node degrees after and before the edge breaking. All stars are shown in blue, while members of the calibration set are in orange. The dashed red line indicates the threshold below which we consider the star as ``migrator.''}
        \label{Fig:breaking_links}
\end{figure}

Thus, given \texttt{model$\_$38} we calculated the ratio R = $\omega$/$\omega_{\rm self}$ for all the edges in the graph. Their cumulative distribution is shown in Fig.~\ref{Fig:breaking_links} together with that of R$_{\rm mem}$. As expected, the comparison between these two curves shows that edges connecting members of the same cluster have retained most of their weights compared to the vast majority of the other edges of the graph. Now, we break all edges with R~$<$~0.2 as they are probably linking stars with strongly dissimilar chemical patterns. Note that by doing so, we remove about 40$\%$ of the total edges. However, only 4$\%$ of the edges between cluster members are broken.

Breaking these edges has changed the number of inward edges of many nodes. In order to identify stars that have possibly migrated, we consider the nodes' degree\footnote{The degree of a node is the number of edges that are linked to that node.} before and after the breaking of non-significative edges. In the bottom panel of Fig.~\ref{Fig:breaking_links} we plot the cumulative histogram of the ratios between the degree after and before the edges breaking. A typical star in our dataset has lost between 30 and 70$\%$ of their original edges with one-$\sigma$ confidence level. We define as ``migrating stars'' all nodes with a ``degree variation rate'' lower than 0.5. These thresholds applied to R and to the degree variation rate -- i.e., 0.2 and 0.5, respectively -- are arbitrary. However, these values are chosen to ensure that approximately half of the stars in the dataset are classified as ``migrating,'' while minimizing the number of stars from the ''calibration set``. In fact, these latter are still gravitationally bound to their siblings, thus they should not be labeled as ``migrators.'' Namely, using the criteria above, we label 25,428 as migrating stars, of which only 7 are cluster members from the calibration set.

\begin{figure*}[t]
\centering
\includegraphics[width=\textwidth]{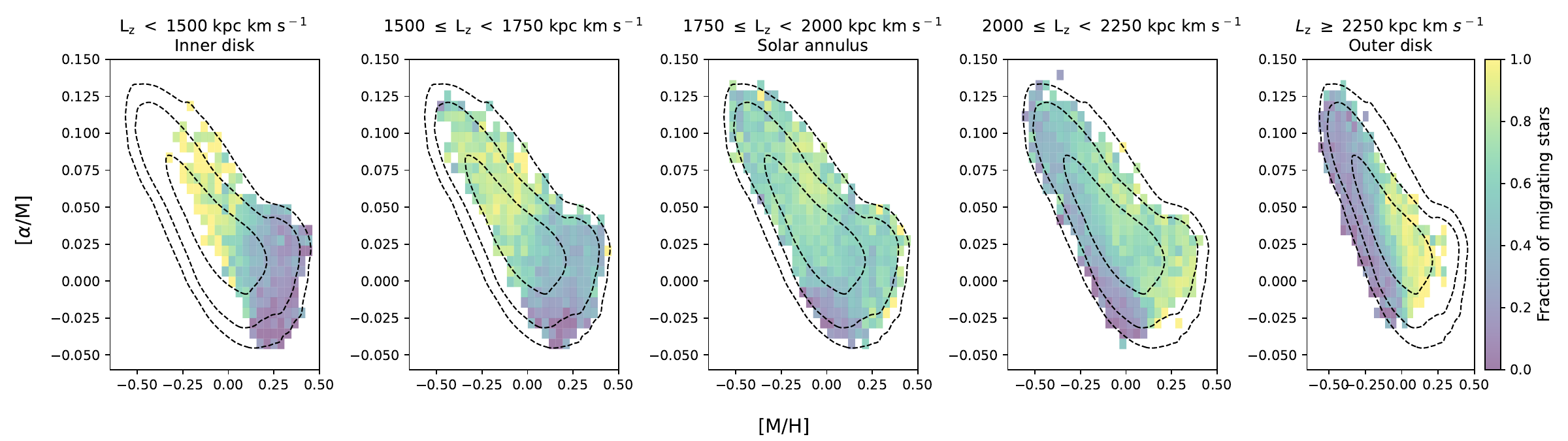}
    \caption{Five panels showing the fraction of migrating stars within the [$\alpha$/M]-[M/H] diagram at different L$_{\rm z}$ bins. The left panel corresponds to the inner disk, and the right panel to the outer disk. The contours represent the distribution of all stars within the diagram at all L$_{\rm z}$.}
        \label{Fig:migrating_stars}
\end{figure*}

The distribution of migrating stars in the [$\alpha$/M]–[M/H] plane is shown in Fig.~\ref{Fig:migrating_stars}. Here the sample is divided into five bins of L${\rm z}$, with the rightmost panel representing stars on orbits closer to the Galactic center, and the leftmost panel corresponding to stars in the outer disk. Within each panel, pixels are colored according to the fraction of migrating stars they contain. For reference, the overlaid contours trace the distribution of the entire stellar sample, regardless of L${\rm z}$.

A significant fraction of stars belonging to the inner disk (leftmost panel) have [M/H] greater than solar. Accordingly to our selection, these stars are mostly formed in situ. There is also a considerable amount of migrating stars which are those with lower [M/H]$<$0 dex. These must have formed in metal-poorer environments, such as the outer disk, and then they migrated inward. The opposite scenario is outlined by the stars currently living in the outer disk (rightmost panel). Most of the stars have metallicities lower than solar. However, the in situ stars are those with lowest [M/H], while the migrating stars have higher [M/H]. These latter have formed in metal-richer environment, thus are probably formed closer to the Galactic centre and then have migrated outward. Instead, migrating stars within the solar annulus are more uniformly distributed across all metallicities. Interestingly, in all these five panels we also observe that the migrating stars have relatively high [$\alpha$/M] values, suggesting that these stars are on average older than the stars formed in situ. These results are consistent with our expectations. In fact, there is now strong evidence that the inner disk is more metal-rich than the outer disk. Although stars with a range of metallicities can be found at any given Galactocentric radius, metal-poor stars in the inner regions must have migrated from the outer disk, while metal-rich stars in the outer regions likely originated in the inner disk. Moreover, stars that have migrated should, on average, be older than stars formed in situ, simply because migration is a process that can take time to occur.

We emphasize that the procedure and the analysis described above in this section, involving the breaking of edges and the analysis of degree variations, is not part of our core methodology for chemical tagging. Rather, it is presented here as an independent validation step aimed at demonstrating that the model has effectively learned to encode chemical similarity/dissimilarity between stars through the edge weights of the graph. In the next section, we resume the main analysis, which focuses on chemical tagging and relies solely on the output of the GATE as generated by \texttt{model$\_$38}.

\subsection{Clustering}
\label{Sec:clustering}

\begin{figure}
\centering
\includegraphics[width=\hsize]{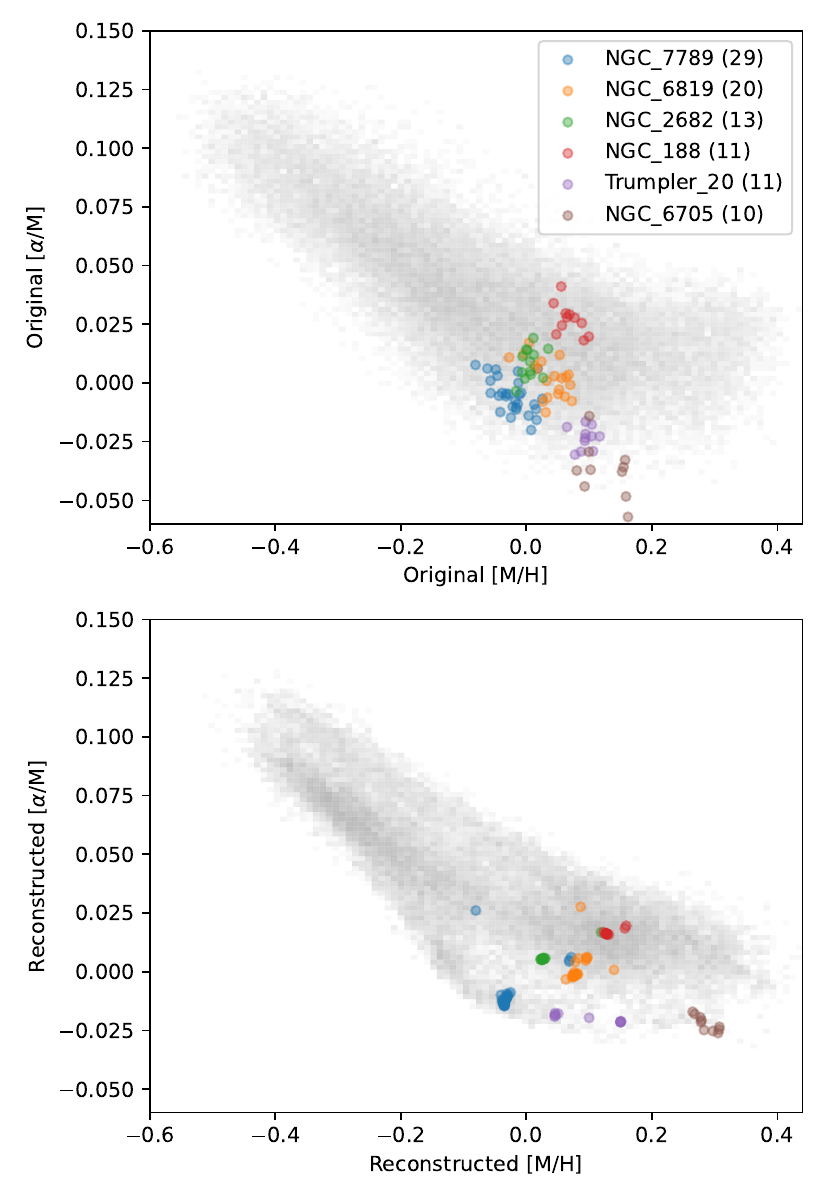}
    \caption{\textbf{Top panel}. The plot shows the distribution of stars from the full dataset (shades of gray) and members of the open clusters from the target set (colored points) in the original [$\alpha$/M]-[M/H] diagram. \textbf{Bottom panel}. As a comparison we show the same diagram of the top panel reconstructed by the GATE. This chemical space has been selectively informed by kinematics as it is described in section~\ref{Sec:Methods}.}
        \label{Fig:abu_diagrams}
\end{figure}

Given that \texttt{model$\_$38} has learnt the information on the chemical similarity or dissimilarity between stars, its output is expected to differ from the input in one key aspect: stars connected by edges and that are indeed chemically similar should overlap even more in the reconstructed chemical space. That happens because the GATE, across all its layers, aggregates information between stars connected by influential edges. The actual effect of this aggregating is seen in Fig.~\ref{Fig:abu_diagrams} where the top panel shows the original [$\alpha$/M]-[M/H] diagram, while the reconstructed one is in the bottom panel. The recontructed [$\alpha$/M]-[M/H] diagram exhibits distinct overdensities and sparse regions that are not present in the original chemical space. 

These artificial features, introduced by GATE, are likely due to differing origins of the stars. For instance, it is expected that stars that have migrated should appear in different regions of the diagram compared to those that have formed in situ, even if they have very similar original chemical patterns. We also observe that Trumpler~20 is split in two high-density clusters, as a consequence of the disconnected-like structure of its subgraph shown in Fig.~\ref{Fig:Trumpler_20_graph}. 

Thus, all of the patterns seen in the bottom panel of Fig.~\ref{Fig:abu_diagrams} likely contain meaningful information; however, interpreting them lies beyond the scope of this paper. Instead, here we focus in observing that members of the target clusters — those we aim to recover as overdensities in chemical space — are significantly more tightly grouped in the reconstructed diagram compared to their distribution in the original one. For that reason, performing clustering analysis in the space reconstructed by GATE is more advantageous than doing so in the original chemical space.

The clustering analysis was applied to the ten features predicted by \texttt{model$\_$38}. For this purpose, we used OPTICS \citep{Ankerst99}, a density-based clustering algorithm. Unlike traditional clustering methods, OPTICS does not produce an explicit clustering of the dataset. Instead, it creates an augmented linear ordering of the data points that reflects their underlying density-based structure. In this ordering, spatially close points appear near each other, and to each point is assigned a ``reachability distance'' -- a measure of the density of its ``neighbouring'' region. The ``reachability plot'' is a visualization of this structure: it displays the reachability distance of each point (on the y axis) as a function of its position in the ordered sequence (on the x axis). In this plot, clusters appear as valleys -- contiguous regions of low reachability distance -- while steep increases in reachability distance typically signal transitions between dense and sparse regions, marking cluster boundaries. Thus, beyond segmenting the dataset into clusters and outliers, the reachability plot serves as a powerful diagnostic tool, offering intuitive insights into the density distribution and intrinsic structure of the data.

The shape of the reachability plot was determined by a single key hyperparameter: \texttt{min$\_$samples}. This parameter controls the size of the neighborhood considered around each point in calculating the reachability distance. When \texttt{min$\_$samples} was set to a low value, neighborhoods are small, making the computed reachability distances highly sensitive to local fluctuations and noise. Conversely, a higher \texttt{min$\_$samples} value leads to larger neighborhoods, which smooths out noise but can also reduce sensitivity to genuine variations in point density. Selecting an appropriate \texttt{min$\_$samples} value is therefore a trade-off: it should be small enough to reflect the true structure and sparsity of the data, but large enough to suppress the influence of noise. 

Before running OPTICS, we sort the dataset on the [M/H] feature, and we standardize, as it is recommended by \citet{Spina22}. Following this, we adopt a commonly used heuristic to set \texttt{min$\_$samples} to approximately twice the number of input features \citep{Ester96}. Given our 10D feature space, we test values of \texttt{min$\_$samples} ranging from ten to 30. We then select the value that minimizes the median distance between adjacent members of the same cluster -- restricted to stars in the calibration set -- within the ordering generated by OPTICS. The clustering analysis is always carried out using the ``manhattan'' metrics, which is also recommended by \citet{Spina22} as it is more stable against noise. Using these criteria, the selected \texttt{min$\_$samples} is 13. 

\begin{figure*}[t]
\centering
\includegraphics[width=\textwidth]{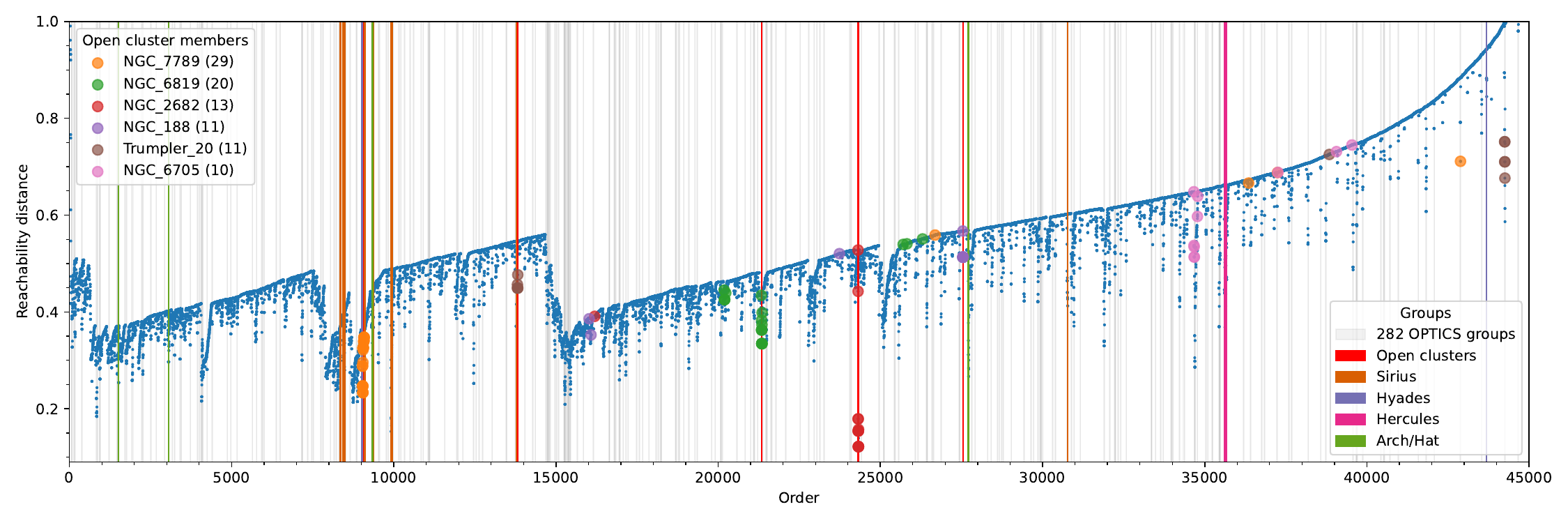}
    \caption{Reachability plot obtained by OPTICS on the chemical space generated by the GATE. Open cluster members from the target set are shown by large colored points. The vertical bands indicate all the groups identified as sharp drops in the reachability distance. The colored vertical bands indicate the groups that actually coincide to known open clusters (red bands) or moving groups.
              }
        \label{Fig:clustering}
\end{figure*}

The resulting reachability plot is shown in Fig.\ref{Fig:clustering}. Several dips are visible: some are deep with a sharp drop in reachability distance, while others are shallower. Most cluster members are located within these features. This is especially clear for the more populous clusters NGC~7789, NGC~6819, and NGC~2682. NGC~188 and NGC~6705 also appear as smaller, concentrated agglomerates. In contrast, Trumpler~20 is split into two high-density groups, mirroring the distribution of its members in the [$\alpha$/M]–[M/H] diagram (Fig.\ref{Fig:abu_diagrams}), which in turn is a consequence of the structure of its graph (Fig.~\ref{Fig:Trumpler_20_graph}).

Among the many dips in the reachability plot, some correspond to genuine overdensities in the reconstructed chemical space, while others result from random density fluctuations caused by noise. A straightforward way to distinguish between the two is to examine the steepness of the reachability distance at the beginning of each dip -- sharp drops typically indicate the onset of a meaningful high-density group. To guide this selection, we employ the \texttt{xi} hyperparameter, which controls the minimum relative decrease in reachability distance required to consider a new high-density group. In essence, \texttt{xi} defines the sensitivity of the clustering algorithm to density changes: smaller values make the algorithm more permissive in identifying high-density groups, while larger values make it more conservative. We test various \texttt{xi} numbers between 0.01 and 0.1 with steps of 0.01 and we evaluate the number of high-density groups with at least more than five stars. A \texttt{xi} = 0.1 extracts 51 groups, which is a quite small number considering the complexity of the reachability plot shown in Fig.~\ref{Fig:clustering}. As expected, lowering \texttt{xi} leads to a steady increase in the number of identified groups, reaching over 1000 when \texttt{xi} = 0.01. Higher \texttt{xi} values tend to isolate only the most prominent overdensities, while smaller values begin to capture finer—and potentially noisy—structures. To balance this trade-off, we select the \texttt{xi} value corresponding to the “knee” of this ``\texttt{xi}-number of groups'' curve, where the number of detected groups begins to grow rapidly with little added significance. This occurs for \texttt{xi} = 0.05, which captures 282 high-density groups with more that five stars.

By definition, these 282 groups are formed by stars clumping in the recontructed chemical abundance space, which is ``informed'' by stellar kinematics and ages. In the next section we show that a fraction of these groups actually coincide with known open clusters and moving groups.

\section{Results}
\label{Sec:results}

The previous section describes the technique used to identify 282 groups of stars -- potentially sharing a common origin -- out of a sample of 47,000 stars. Here, we show that some of these groups indeed correspond to real stellar associations that we aim at recovering. Others, however, are likely to be spurious clumps -- overdensities identified in the reconstructed chemical space, possibly due to the strong non-uniformity of APOGEE’s selection function or other noise-related effects.

One could simply review all these groups and analyze the distribution of their stars in specific diagnostic diagrams to identify potentially interesting cases. This is feasible for 282 groups, as the number is still manageable. However, it is also possible to define automated methods to further narrow down the selection to a smaller subset of the most promising groups. The specific criteria would depend on the type of structures we target. For example, both open clusters and moving groups are expected to be tightly clustered in orbital action space. Therefore, one possible strategy would be to focus the analysis on groups that meet this compactness criterion.

To identify stellar groups that are more dynamically coherent in orbital action space, we computed for each of these groups the median absolute deviation (MAD) in L$_{\rm Z}$, J$_{\rm Z}$, and J$_{\rm R}$. This yields a single 3D vector per group, summarizing its internal dynamical spread in a robust, outlier-resistant way. We then model the distribution of all groups in this MAD space using a Gaussian multivariate copula. The copula captures the dependence structure among the MADs across different action dimensions, enabling a multivariate assessment of group compactness. Using the fitted copula, we compute the cumulative distribution function (CDF) for each group’s MAD vector. This CDF value reflects how statistically ``extreme'' the group’s dispersion is, relative to the full population of groups. Finally, we select the most compact groups as those with CDF values below a given threshold (e.g., CDF~$<$~0.01). These groups that are simultaneously compact in all three action dimensions. 

\subsection{Open clusters}
\label{Sec:open_clusters}

The target set defined in section~\ref{Sec:Dataset_processing} is composed by the six open clusters that in our dataset have ten or more members: NGC~7789, NGC~6819, NGC~2682, NGC~188, NGC~6705, Trumpler~20. These are the stars we are ultimately aiming to recover as independent groups in order to validate our method of ``chemical tagging.'' 

Since open clusters are expected to be highly compact in orbital parameter space, we consider only those among the 282 groups with a CDF value less than 0.01. This reduces the sample to 12 groups, which include five of the six clusters from the target set.

\begin{figure*}[t]
\centering
\includegraphics[width=\textwidth]{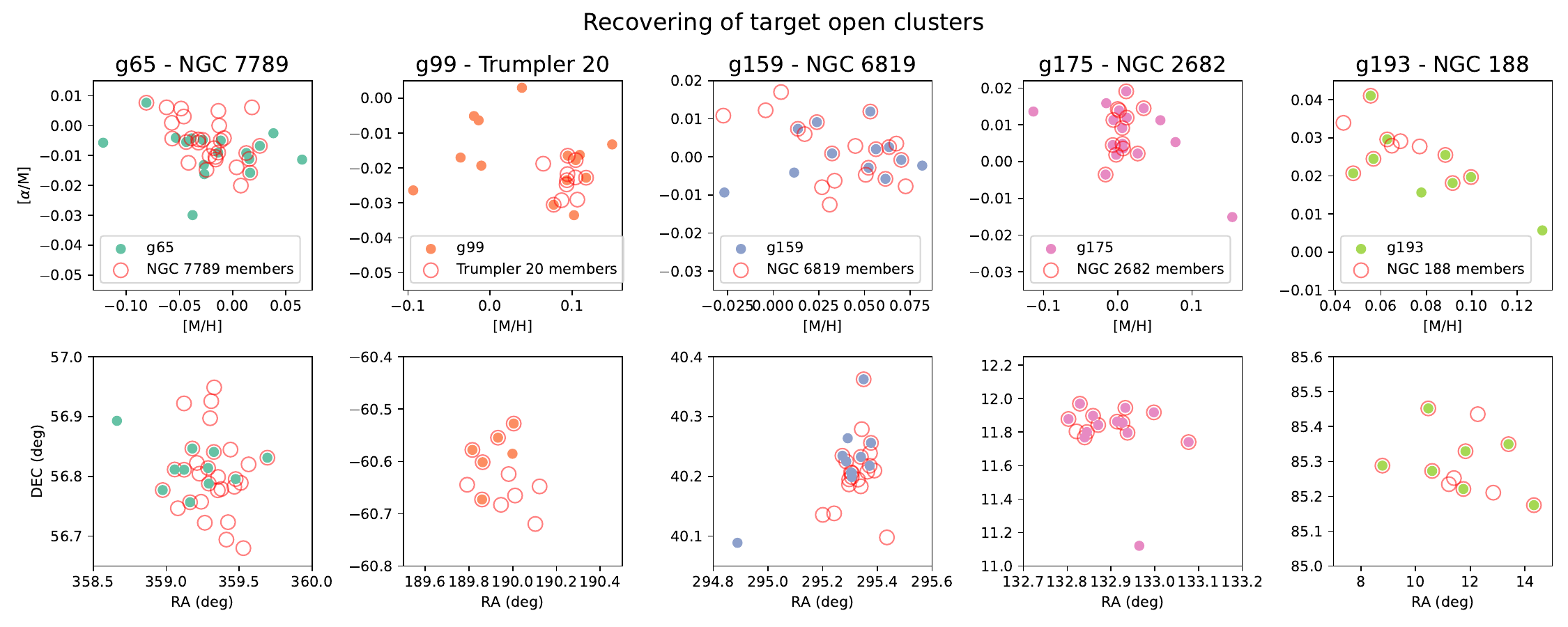}
    \caption{These panels illustrate the effectiveness of our method in recovering the target open clusters. The top panels display, for each group and its associated open cluster, the distribution of stars in the [$\alpha$/M]–[M/H] plane. Filled symbols represent the stars belonging to each high-density group identified by OPTICS, while red circles indicate the true members of the open clusters. The bottom panels provide a zoomed-in view centered on the cluster sky coordinates.}
        \label{Fig:recovered_clusters}
\end{figure*}

These groups and their corresponding open clusters are shown in Fig.\ref{Fig:recovered_clusters}, while statistical parameters that quantify the retrieval performance are listed in Table~\ref{tab:tabClusters}. For instance, \textit{g175} is one of the groups that most effectively recovered a target cluster. This group exhibits a homogeneity of 71$\%$, meaning that 12 out of its 17 stars belong to NGC~2682 (i.e., 12/17 = 0.71), one of the target clusters. The completeness of this recovery is 92$\%$, indicating that 12 of the 13 known members of NGC~2682 are included in \textit{g175} (12/13 = 0.92). Each of the five recovered target clusters is associated with a group that achieves both homogeneity and completeness above 30$\%$.

To assess the robustness of these results with respect to observational uncertainties, we repeated the graph construction and clustering analysis across ten independent realizations of the input data. In each case, orbital parameters were re-derived from Monte Carlo sampling of astrometric and radial velocity uncertainties. We find that $\sim$75$\%$ of graph’s edges are the same of the original graph and that a median of four open clusters are recovered with completeness and homogeneity above 0.3. This confirms that our method is resilient to typical observational perturbations. Moreover, the large sample size of over 47,000 stars ensures that the impact of individual uncertainties is statistically diluted, further enhancing the robustness of the identified structures.

The performance of our method can be compared with two of the most recent studies on chemical tagging in the thin disk. For instance, \citet{Casamiquela21} analyzed a sample of 175 stars and successfully recovered approximately $\sim$30$\%$ of the open clusters present in their dataset. Similarly, \citet{Spina22} achieved a recovery rate of about  $\sim$50$\%$ using a sample of 568 stars. In contrast, our method identifies 5 out of 6 open clusters within a much larger sample of 47,000 stars.

The top row of Fig.~\ref{Fig:recovered_clusters} shows the original [$\alpha$/M]-[M/H] diagrams of groups and the open clusters they include. More specifically, filled points represent stars belonging to a given group, while the red empty circles are the actual members of the open cluster. Here, it is possible to distinguish the cluster members that have been successfully recovered from the ``contaminants,'' which are all the other stars within the group. It is also possible to identify which cluster members are missing from the group. In general, most contaminants are located at the outskirts of the cluster member distributions. This suggests that the group is slightly more chemically extended than the cluster itself, at least in the [$\alpha$/M]–[M/H] plane (e.g., see g99 and g175). However, it is important to keep in mind that this projection shows only two out of the ten dimensions used in the clustering analysis. In the full high-dimensional space, these apparent outliers may actually be consistent with the cluster distribution -- what appears as an outlier in two dimensions may be an inlier in others. In our method, all chemical dimensions are treated with equal weight, but an alternative approach could prioritize elements with typically higher precision, such as [M/H] and [$\alpha$/M], potentially leading to different group definitions.

The bottom row of Fig.~\ref{Fig:recovered_clusters} shows the zoomed-in view around the cluster's sky coordinates. In these plots we notice that some contaminants are located very close to the cluster members. These stars are actually members of the clusters (see \citealt{Hunt23}), but either are outside the tidal radius or they have a membership probability lower than 0.5. Hence, in section~\ref{Sec:Dataset_processing} they were not included among the sample of cluster members, although they probably are. 

\begin{table*}[h]
    \centering
    \small
\caption{Statistical parameters quantifying the open clusters' retrieval performances.}
\begin{tabularx}{\textwidth}{X X X X X}
 \hline
 \hline
$\bf{Group~65~(NGC~7789)}$ & $\bf{Group~99~(Trumpler~20)}$ & $\bf{Group~159~(NGC~6819)}$ & $\bf{Group~175~(NGC~2682)}$ & $\bf{Group~193~(NGC~188)}$ \\ \hline
Group size: 17 & Group size: 14 & Group size: 12 & Group size: 17 & Group size: 9 \\
All cluster members: 29 & All cluster members: 11 & All cluster members: 20 & All cluster members: 13 & All cluster members: 11 \\
Retrieved memb.: 10 (+1) & Retrieved memb.: 5 (+1) & Retrieved memb.: 9 (+2) & Retrieved memb.: 12 (+1) & Retrieved memb.: 7 \\
Completeness: 0.34 & Completeness: 0.45 & Completeness: 0.45 & Completeness: 0.92 & Completeness: 0.64 \\
Homogeneity: 0.59 & Homogeneity: 0.36 & Homogeneity: 0.75 & Homogeneity: 0.71 & Homogeneity: 0.78 \\ \hline

\end{tabularx}
    \label{tab:tabClusters}
\end{table*}

Besides the groups discussed above, among the 12 groups there is one corresponding to the open cluster NGC~8611, which is part of the ``training set.'' As we demonstrate in the next paragraphs, most of the remaining groups are other stellar associations of interest.

\subsection{Moving groups}
\label{Sec:moving_groups}

\begin{figure*}[t]
\centering
\includegraphics[width=\textwidth]{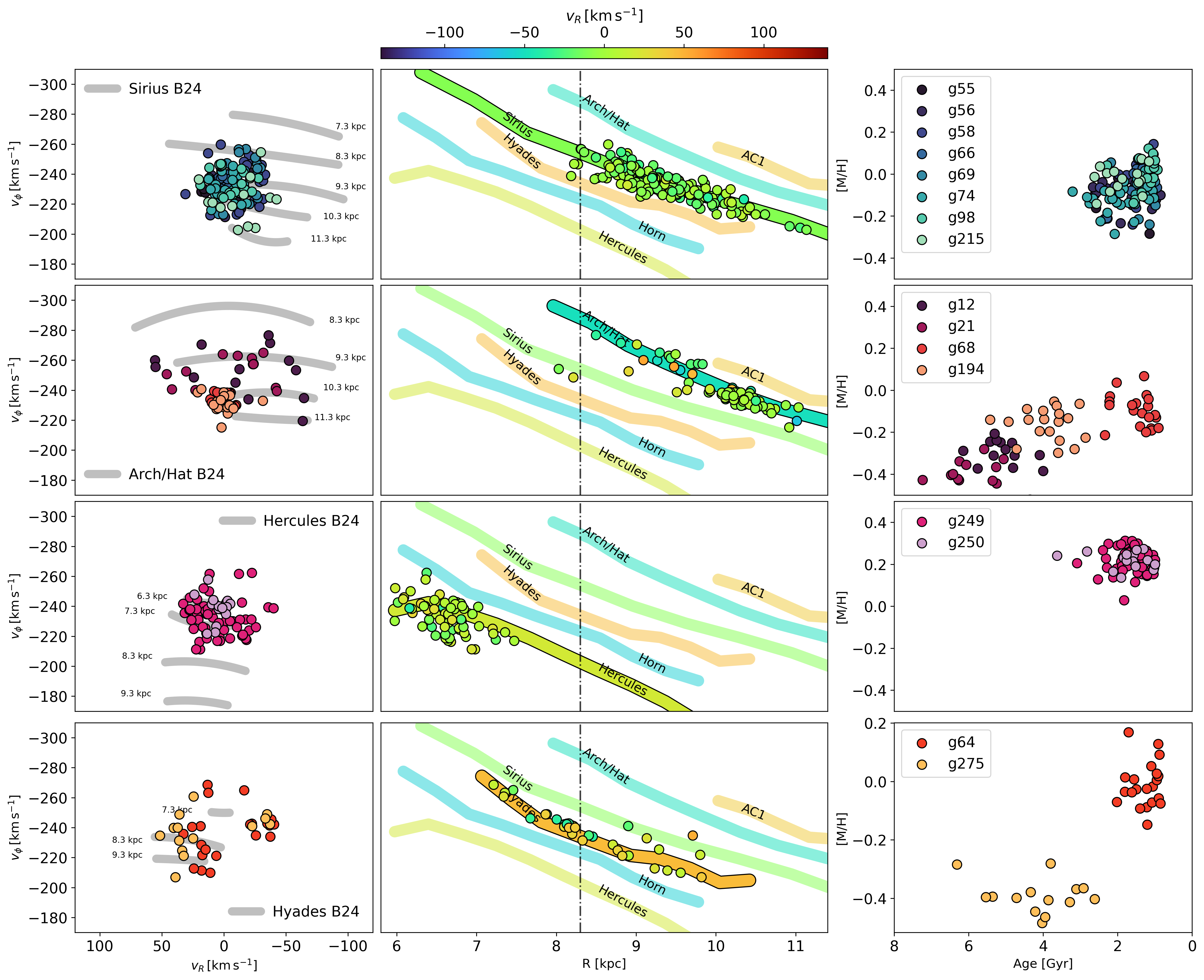}
    \caption{Overview of the high-density stellar groups in relation to known moving groups. Each row corresponds to one moving group recovered through our method. From top to bottom we show Sirius, Arch/Hat, Hyades, and Hercules. Left and middle panels show distributions of stars in the v$_{\phi}$–v$_{\rm R}$ and v$_{\phi}$–R$_{\rm Gal}$ diagnostic diagrams, respectively. Points are color-coded by high-density group (left) and by v$_{\rm R}$ (middle), with overlaid regions indicating the locations where overdensities were identified by \citet{Bernet24}. The arches and ridges reflect how the moving groups vary with R$_{\rm Gal}$ and v$_{\rm R}$. Right panels show the distribution of stars in the [M/H]–Age plane, color-coded by group.}
        \label{Fig:groups_all}
\end{figure*}

Galactic moving groups are collections of stars that share similar space velocities, despite being spread over several Galactocentric radii. They appear as overdensities in Galactic velocity space, particularly in the v$_{\phi}$-v$_{\rm R}$ and v$_{\phi}$-R$_{\rm Gal}$ diagrams, where v$_{\phi}$ and v$_{\rm R}$ represent the azimuthal and radial velocities in the Galaxy’s cylindrical coordinate system, and R$_{\rm Gal}$ denotes the Galactocentric radius \citep[e.g.,][]{Ramos18,Bernet22,Lucchini23,Bernet24}. These peculiarities likely indicate a common nature or dynamical history. The possible origins of these moving groups include trapping by bar resonances, trapping by spiral arm resonances, perturbation from transient spiral structure, or phase mixing following some perturbation \citep[e.g.,][]{Hunt18,Antoja22,Lucchini24}. Although stars in a moving group are not necessarily chemically homogeneous -- as is typically the case for stars in an open cluster -- it is still possible that stars formed from the same material, and therefore similar in chemical composition, have been captured by these dynamical structures. As a result, moving groups may indeed consist of one or more subgroups of chemically similar stars. If this is the case, then we expect to identify some of these dynamical populations among the groups identified through our method.

We expect moving groups to be relatively compact in orbital action space, although not as tightly clustered as open clusters, which are gravitationally bound systems. Therefore, among the 282 groups identified with OPTICS, we focus on those with a CDF~$<$~0.1 and containing at least ten stars. This selection narrows the sample to 56 high-density groups of potential interest. The distribution of stars in 19 of these groups within the diagnostic v$_{\phi}$-v$_{\rm R}$ and v$_{\phi}$-R$_{\rm Gal}$ diagrams aligns with the locations of known moving groups such as Arch/Hat, Sirius, Hyades, and Hercules. 

Figure~\ref{Fig:groups_all} shows how stars from the 19 high-density groups align with the positions of these moving groups. Each row corresponds to a single moving group. The left and middle panels in each row display the distribution of stars in the diagnostic diagrams v$_{\phi}$–v$_{\rm R}$ and v$_{\phi}$–R$_{\rm Gal}$, respectively. In the v$_{\phi}$–v$_{\rm R}$ diagrams, points are color-coded according to their high-density group, while in the v$_{\phi}$–R$_{\rm Gal}$ diagrams, they are colored based on their v$_{\rm R}$. This color-coding allows the reader to easily track each point and its corresponding group across both diagrams.

These two diagnostic plots also include overlays of the regions where the specific moving groups were identified as overdensities by \citet{Bernet24}. The locations of moving groups in the v$_{\phi}$–v$_{\rm R}$ diagram typically vary with R$_{\rm Gal}$ (see, for example, Fig. 1 in \citealt{Bernet22}). For this reason, we display different arches at different Galactocentric distances. Similarly, in the v$_{\phi}$–R$_{\rm Gal}$ diagrams, we highlight the ridges where moving groups are generally found. Although stars in a moving group can span a wide range of v$_{\rm R}$ -- a detail not directly visible in the v$_{\phi}$–R$_{\rm Gal}$ diagrams -- the slope and extent of these ridges can vary with v$_{\rm R}$ (see Fig. 2 in \citealt{Bernet22}). In our plots, we color each ridge according to the v$_{\rm R}$ that corresponds to the longest extent in R$_{\rm Gal}$. While this does not mean all stars in a given moving group have that specific v$_{\rm R}$, we do observe a general agreement between the stars’ v$_{\rm R}$ values and those of the ridge with the largest R$_{\rm Gal}$ span.

Finally, the right panels of Fig.~\ref{Fig:groups_all} show the distribution of stars from the high-density groups in the [M/H] versus Age plane. As in the left panel, points are color-coded according to their group. Most of the groups identified through our method are predominantly composed of young stars. Overall, the metallicities of these young stars are consistent with those reported by \citet{Lehmann25} for the same moving groups. In the cases of the Arch/Hat and Sirius groups, we also detect stars older than 2 Gyr with lower metallicities compared to their younger counterparts. The distribution of stars in these two groups traces a sequence from older, more metal-poor stars to younger, more metal-rich ones, likely reflecting the chemical evolution within the Galactic disk. Furthermore, we find that chemical enrichment in the Sirius group has occurred at a faster rate than in the Arch/Hat group. This is expected, given that Sirius is located closer to the Galactic center, where chemical evolution proceeds more rapidly.

\subsection{Clustering in the original chemical space and the latent space}

To evaluate the effectiveness of the GATE-reconstructed chemical space, we compared the clustering performance against the original chemical abundance space. Specifically, we applied the same clustering procedure used in the reconstructed space: OPTICS, with the same hyperparameter configuration, directly to the original 10D chemical abundance space.

The results show a stark contrast: in the original chemical space, OPTICS fails to recover any of the six open clusters in the target set. This outcome confirms previous findings in the literature that the chemical homogeneity of the thin disk poses a fundamental limitation to chemical tagging based solely on abundance information \citep[e.g.,][]{BlancoCuaresma15, Casamiquela21, Spina22}.

As an additional benchmark, we also apply the clustering analysis to the latent space (i.e., the 4D bottleneck representation learned by the GATE). There, we are able to recover only three open clusters (NGC 7789, Trumpler 20, and NGC 2682) with homogeneity and completeness above 0.3. The reduced performance in the latent space compared to those of the reconstructed 10D chemical space can be attributed to two key factors. First, the latent space -- which represents a compressed abstraction of the input data -- often loses fine chemical details that are crucial for distinguishing co-natal groups, especially in the chemically homogeneous thin disk. Second, autoencoder's latent space is not regularized to be continuous or well-structured for clustering \citep[e.g.,][]{Song13}, even when the autoencoder achieves excellent reconstruction performance.

Together, these comparisons underscore the importance of our chemically informed representation. While both the original chemical space and the latent space fall short in reliably identifying co-natal groups, the GATE-reconstructed chemical space achieves significantly improved clustering performance.

\section{Conclusions}
\label{Sec:conclusions}

One primary limitation of chemical tagging within the thin disk is the high level of chemical homogeneity of the stars. That reduces the chemical contrast between co-natal and unrelated stars, hindering the robust reconstruction of disrupted stellar populations. Thus, the effectiveness of chemical tagging is critically linked to the precision of chemical abundance measurements. Current and upcoming spectroscopic surveys do not yet provide the necessary abundance precision for fully effective chemical tagging.

To overcome these limitations, we developed and tested a new analysis method. Our approach remains fundamentally anchored in chemistry, but is designed to leverage kinematic and age information in a synergistic and efficient way. The method is built upon a graph attention auto-encoder (GATE). In this framework, stars are represented as nodes in a graph, with their chemical abundances as node features. Edges are established between stars with similar orbital actions and ages. The GATE learns the importance (i.e., weight) of these connections for reconstructing the stellar chemical patterns. Hence, information on kinematics and age is effectively exploited only when it results to be chemically relevant, which makes the full methodology chemically grounded. The output of the trained GATE provides an ``informed'' representation of the chemical abundance space, where stars connected by influential edges are drawn closer together. This process results in a overdensities of such stars that can be identified through clustering analysis.

The new method was applied to a sample of $\sim$47,000 thin disk stars and a feature space of ten chemical abundances derived by APOGEE. The primary objective of this analysis is the blind recovering of six open clusters composed by ten to 29 stars hidden within this dataset. Our analysis yields several key results:

\begin{itemize}
\item In Fig.~\ref{Fig:abu_diagrams} we show that members of open clusters are significantly more tightly grouped in the GATE-reconstructed space than in the original chemical space. 

\item As a validation step, we show that the edge weights learnt by the GATE correctly capture information on chemical similarity. This information extracted from data can be used to potentially identify stars that have undergone significant orbital changes (see Fig.~\ref{Fig:migrating_stars}). While the primary focus of this work is chemical tagging, this demonstrates the GATE's ability to encode information relevant to stellar history in a fully data-driven fashion.

\item The clustering analysis applied on the GATE-reconstructed space identifies 282 high-density groups of stars potentially sharing a common origin. About $\sim$10$\%$ of these correspond to known stellar populations.

\item By focusing on groups that are compact in orbital action space, we successfully recovered five out of the six target open clusters with ten or more members. For these recovered clusters, homogeneity and completeness metrics were above 30$\%$, with one group achieving 92$\%$ completeness and 71$\%$ homogeneity for NGC 2682 (see Fig.~\ref{Fig:clustering}).

\item Expanding our search to groups with slightly less orbital compactness (CDF < 0.1) and at least ten stars, we found that the distribution of stars in 19 of the identified high-density groups aligns with the locations of known moving groups such as Arch/Hat, Sirius, Hyades, and Hercules in diagnostic velocity and radial distance diagrams (see Fig.~\ref{Fig:groups_all}).

\end{itemize}

In summary, our GATE-based method offers a promising new pathway for improving the ability to identify dispersed stellar groups in the thin disk. By selectively informing the chemical space with kinematic and age information, the method preserves the chemical basis of tagging while allowing for a more effective identification of co-natal populations and dynamical associations like moving groups, even in the presence of challenging chemical homogeneity and observational uncertainties.

\begin{acknowledgements}
      We thank Teresa Antoja for the interesting discussion and Marcel Bernet for sharing with us data from their recent works.
\end{acknowledgements}

\bibliographystyle{aa}
\bibliography{Bibliography.bib}

\end{document}